\documentclass[12pt]{article}
\usepackage{amsmath}

\begin{document}
\title {Supersymmetry and Entanglement in the Generalized Dirac Oscillator}

\author{
H. P. Laba$^1$,
V. M. Tkachuk$^2$\footnote{voltkachuk@gmail.com}\\
$^1$Department of Applied Physics and Nanomaterials Science, \\
Lviv Polytechnic National University,\\
5 Ustiyanovych St., 79013 Lviv, Ukraine,\\
$^2$Department for Theoretical Physics,\\
Ivan Franko National University of Lviv,\\
12, Drahomanov St., Lviv, 79005, Ukraine.}

\maketitle

\begin{abstract}
The supersymmetric properties of the generalized Dirac oscillator allow us to determine the entanglement entropy between the spin and the continuous variable in a purely algebraic manner. The entanglement has a relativistic origin and disappears in the nonrelativistic limit. The entanglement entropy attains its maximal value in the limit of infinite energy.



\end{abstract}

\section{Introduction}

Among relativistic quantum models, the Dirac oscillator occupies a special place as one of the few systems that admit exact solutions. It was first introduced in \cite{Ito67} as a relativistic model with a Hamiltonian linear in both momentum and coordinate operators. Later, the system was reintroduced in \cite{Mosh89}, where it was given the name Dirac oscillator. This work stimulated extensive research activity and established the Dirac oscillator as an important model in relativistic quantum mechanics. An experimental realization of the Dirac oscillator was subsequently proposed in \cite{Fra13}. A concise historical overview of the development of the Dirac oscillator can be found in \cite{Que17}.

The Dirac oscillator possesses a natural supersymmetric (SUSY) structure since its equations can be factorized in terms of first-order operators, leading to a pair of partner Hamiltonians in the sense of Witten's 
supersymmetric quantum mechanics \cite{Coo95,Gan11}. 
This connection enables the application of shape invariance, factorization methods, and algebraic techniques developed in supersymmetric quantum mechanics to obtain exact and quasi-exact solutions for Dirac systems \cite{Ho2004,Ikh12,Dut13,Junker2020,Nat14,LabaTka18,Wei2021}. Note that the Dirac oscillator can be solved exactly in the case
of quantized space with minimal length \cite{Que05,Que06}.
These studies demonstrate that supersymmetry provides a unifying framework for understanding exact solvability, spectral degeneracies, and algebraic properties of generalized Dirac oscillator systems.


Relativistic quantum systems described by the Dirac equation possess intrinsic entanglement between spin and other degrees of freedom, such as momentum, position, or particle--antiparticle components. For a free Dirac particle, the Hilbert space naturally factorizes into spin and momentum subspaces, allowing the investigation of spin--momentum entanglement and its transformation under Lorentz boosts. It has been shown that relativistic transformations generally modify the amount of spin entanglement due to Wigner rotations, making entanglement observer-dependent \cite{Peres2002,Gingrich2002}. Furthermore, the Dirac bispinor structure itself can be interpreted as a composite system consisting of spin and intrinsic parity qubits, leading to studies of quantum correlations and entanglement in relativistic fermionic states \cite{Caban2005,Debarba2017}.

The Dirac oscillator provides a particularly important relativistic model for entanglement studies. Owing to its exact mapping onto the Jaynes--Cummings and anti--Jaynes--Cummings models, the Dirac oscillator naturally generates entanglement between discrete spin degrees of freedom and continuous oscillator variables \cite{Bermudez2007}. Entanglement dynamics, quantum coherence, and nonclassical correlations have been investigated in one- and two-dimensional Dirac oscillators, as well as in the presence of external fields \cite{Villalba2008,Bermudez2008}. The model has also attracted considerable interest in quantum simulation, where relativistic effects such as Zitterbewegung, Klein tunneling, and spin--orbital entanglement can be explored in controllable quantum platforms \cite{Lamata2007}. Consequently, the Dirac oscillator serves as a bridge between relativistic quantum mechanics, quantum optics, and quantum information theory.


Interest in quantum entanglement in relativistic systems has continued to grow in recent years. A comprehensive modern review of relativistic spin and entanglement was presented in \cite{Land2024}, where the covariant description of spin and quantum correlations in relativistic quantum mechanics was presented. More recently, the authors of paper \cite{Chargui2025} analyzed spin--rest entanglement in solutions of the Dirac equation interacting with a quantized electromagnetic field and a confining potential. Furthermore, the dynamics of entanglement in a time-dependent Dirac oscillator were investigated in Ref.~\cite{Andrade2026}, where entanglement between chirality and oscillator degrees of freedom was quantified using negativity and von Neumann entropy. These works illustrate the growing interplay between relativistic quantum mechanics, quantum information theory, and quantum simulation platforms.


Supersymmetric quantum mechanics (SUSY QM) provides a natural framework where quantum entanglement emerges as a consequence of the graded structure of the Hilbert space. In SUSY systems, bosonic and fermionic sectors are connected by supercharges, which generate transitions between supersymmetric partner Hamiltonians.  SUSY eigenstates can appear as superpositions of bosonic and fermionic components, giving rise to intrinsic entanglement between these sectors. In this sense, SUSY introduces a natural bipartition of the Hilbert space analogous to a 
qubit-oscillator or spin-boson decomposition.
A central result is that energy eigenstates of SUSY QM with indefinite fermion number are generally entangled states. This was explicitly demonstrated in the context of generalized Jaynes-Cummings-type models, where supersymmetry provides a natural mechanism for generating and classifying entangled states \cite{Cattaruzza2014,LabaTkachuk20}. Within this framework, entanglement is closely related to the action of supercharges and the degeneracy structure of SUSY partner spectra. 
The algebraic and information-theoretic approaches have clarified the structure of entanglement in SUSY systems. Using modular operator techniques in von Neumann algebras, it has been shown that entanglement measures such as concurrence can be directly related to algebraic structures generated by SUSY transformations \cite{Chatterjee2021} (see also \cite{LabaTkachuk20,TkaVak08}). In addition, entanglement dualities have been derived for quadratic SUSY Hamiltonians, establishing precise relations between bosonic and fermionic entanglement spectra and revealing that SUSY constraints can strongly shape the scaling and structure of entanglement entropy \cite{Jonsson2021}. These results indicate that SUSY not only determines spectral properties but also imposes nontrivial constraints on quantum correlations.

Despite these developments, the interplay between supersymmetry and quantum entanglement in Dirac oscillator systems remains largely unexplored. Existing studies mainly focus either on the supersymmetric properties of Dirac oscillators or on their entanglement characteristics from the perspective of quantum information theory. To the best of our knowledge, there is no systematic investigation of how supersymmetry and SUSY partner structures affect entanglement measures and quantum correlations in generalized Dirac oscillator models. Consequently, the simultaneous study of supersymmetry and entanglement in generalized Dirac oscillators represents an open problem at the intersection of relativistic quantum mechanics, supersymmetric quantum mechanics, and quantum information theory. Addressing this problem may provide new insights into the role of hidden symmetries in the generation and control of quantum correlations in relativistic systems.
This paper aims to fill this gap.

\section{Generalized Dirac oscillator  and SUSY quantum mechanics}

The eigenvalue equation of the (1+1)-dimensional Dirac oscillator is given by
\begin{eqnarray}
(c\sigma_x(p_x-im\omega x\sigma_z)+mc^2\sigma_z)|\psi\rangle=E|\psi\rangle
\end{eqnarray}
where we set the Planck constant $\hbar=1$ and the momentum operator $p_x=-id/dx$,
$\sigma_{\alpha}$ ($\alpha=x,y,z$) are Pauli matrices, $|\psi\rangle$ is two component wavefunction with large $\phi$ and small $\chi$ components, respectively.

The generalized Dirac oscillator can be obtained from the ordinary Dirac oscillator by making the substitution
\begin{eqnarray}
m\omega x\to W(x),
\end{eqnarray}
where $W(x)$ is a function of the position $x$. 

The eigenvalue equation for the generalized Dirac oscillator
\begin{eqnarray}
(c\sigma_x(p_x-iW(x)\sigma_z)+mc^2\sigma_z)\psi=E\psi
\end{eqnarray}
in explicit form reads
\begin{eqnarray} \label{Eqphi}
mc^2\phi+c\hat B^+\chi=E\phi \\
c\hat B\phi- mc^2\chi=E\chi. \label{Eqchi}
\end{eqnarray}
Here, we introduce the annihilation and creation operators within the framework of supersymmetric (SUSY) quantum mechanics
\begin{eqnarray}\label{BB}
\hat B=\hat p-iW(x), \ \  \hat B^+=\hat p+iW(x),
\end{eqnarray}
$W(x)$ plays the role of superpotential. For a review of SUSY quantum mechanics, see for instance, \cite{Coo95,Gan11}.

From (\ref{Eqchi}) for small component we find
\begin{eqnarray}\label{cp}
\chi={1\over E/c+mc}\hat B\phi.
\end{eqnarray}
Substituting it into (\ref{Eqphi}), we obtain the eigenvalue equation for the large component in the standard form of SUSY quantum mechanics
\begin{eqnarray}\label{B+B}
\hat B^+\hat B\phi=\epsilon\phi,
\end{eqnarray}
where we introduce the notation 
$\epsilon=(E/c)^2-m^2c^2$. 
Explicitly, this equation reads
\begin{eqnarray}\label{SUSY}
(p_x^2+W^2- W')\phi=\epsilon\phi,
\end{eqnarray}
$W'$ denotes derivative with respect to x, ${d W/ dx}$.

Similarly from (\ref{Eqphi}) for large component we find
\begin{eqnarray}\label{pc}
\phi={1\over E/c-mc}\hat B^+\chi
\end{eqnarray}
and equation for $\chi$ reads
\begin{eqnarray}\label{BB+}
\hat B\hat B^+\chi=\epsilon\chi,
\end{eqnarray}
or
\begin{eqnarray}\label{SUSY1}
(p_x^2+W^2+W')\chi=\epsilon\chi,
\end{eqnarray}

Hamiltonians $H_-=\hat B^+\hat B$ and $ H _ {+} = \hat {B} \hat {B} ^+ $ are two SUSY partners in SUSY quantum mechanics. They possess identical energy spectra except for the ground state.
In the case of exact SUSY, only one of them has a zero-energy level $\epsilon=0$ ($E=\pm mc^2$). Let it be $ H _-$.  Then the corresponding wave function satisfies the equation
\begin{eqnarray}
\hat B \phi_0=0
\end{eqnarray}
the solution of which is
\begin{eqnarray}
\phi_0=Ce^{-\int dx W(x)}.
\end{eqnarray}
The superpotential in this case satisfies the condition $W(x)\to \pm {\rm const}$ at 
$x \to \pm \infty$. With these conditions, the function is square integrable.

For a small component, according to (\ref{cp}), we find $\chi_0 = 0$ and ground-state solution with $\epsilon=0$ ($E=mc^2$) reads
\begin{eqnarray}\label{psi0}
|\psi_0\rangle=
\left(
  \begin{array}{c}
  \phi_0\\
    0 \\
  \end{array}
\right)=
|\uparrow\rangle \phi_0(x).
\end{eqnarray} 
Note that the spin degree of freedom and the continuous variable for the ground state are separated.


The excited state of the Dirac oscillator with nonzero $\epsilon >0$ ($E>mc^2$) in general reads
\begin{eqnarray}\label{psi1}
|\psi\rangle=
\left(
  \begin{array}{c}
  \phi\\
    \chi\\
  \end{array}
\right)=
|\uparrow\rangle \phi(x)+|\downarrow\rangle \chi(x),
\end{eqnarray}
where $\phi$ and $\chi$ are SUSY partners and are related by (\ref{cp}) and (\ref{pc}). From normalization condition for $|\psi\rangle$ we have
\begin{eqnarray}
\int_{-\infty}^{\infty}dx (|\phi(x)|^2+|\chi(x)|^2)=1.
\end{eqnarray}

\section{Entanglement of spin and continuous variable}
To calculate the entanglement entropy, it is convenient to rewrite  (\ref{psi1}) in terms of the normalized components of the state vector
\begin{eqnarray}
 \phi(x)=a\tilde\phi(x),  \  \  \  \chi(x)=b\tilde\chi(x),
\end{eqnarray}
where $\tilde\phi(x)$ and $\tilde\chi(x)$ are normalized large and small components of state vector
\begin{eqnarray}
\int_{-\infty}^{\infty}dx |\tilde\phi(x)|^2=1, \ \ 
\int_{-\infty}^{\infty}dx|\tilde\chi(x)|^2)=1.
\end{eqnarray}

Note that $\tilde\phi(x)$ and $\tilde\chi(x)$ are related by SUSY transformation
\begin{eqnarray}
\hat B\tilde\phi(x)=\sqrt \epsilon\tilde\chi(x), \ \ 
\hat B^+\tilde\chi(x)=\sqrt \epsilon \tilde\phi(x).
\end{eqnarray}

It can be obtained from (\ref{B+B}) and (\ref{BB+}) by applying $\hat B$ and $\hat B^+$ to these equations, respectively.

Then, according to (\ref{cp}) and using the SUSY transformation, we find
\begin{eqnarray}
\chi(x)={a\over E/c+mc}\hat B \tilde\phi(x)=
{a\sqrt\epsilon\over E/c+mc}\tilde\chi(x),
\end{eqnarray}
and thus 
\begin{eqnarray}
b={\sqrt\epsilon\over E/c+mc}a=\gamma a
\end{eqnarray}
where we introduce the notation 
\begin{eqnarray}
\gamma={\sqrt\epsilon\over E/c+mc}={\sqrt {(E/c)^2-m^2c^2}\over E/c+mc}=
\sqrt {{E/c-mc}\over {E/c+mc}}.
\end{eqnarray}
One can see that $ 0  \le \gamma \le 1$, where $\gamma=0$ at $E=mc^2$ and 
$\gamma=1$ at $E\to\infty$.

Now we rewrite the excited state in the form
\begin{eqnarray}\label{psi1}
|\psi\rangle=
a|\uparrow\rangle \tilde\phi(x)+b|\downarrow\rangle \tilde\chi(x)=\\
a\left(|\uparrow\rangle \tilde\phi(x)+
\gamma|\downarrow\rangle \tilde\chi(x)\right).
\end{eqnarray}
From the normalization condition $a^2+b^2=1$, we find
\begin{eqnarray}
a={1\over \sqrt{1+\gamma^2}}, \ \
b={\gamma\over \sqrt{1+\gamma^2}},
\end{eqnarray}


Let us consider the general case of an arbitrary superpotential $W(x)$. In this case, the functions $\tilde{\phi}(x)$ and $\tilde{\chi}(x)$ are not necessarily orthogonal.
\begin{eqnarray}
\int_{-\infty}^{\infty}dx \tilde\phi^*(x)\tilde\chi(x)=Z=e^{i\alpha}|Z|,
\end{eqnarray}
$Z$ is the overlap integral between two SUSY partner eigenstates. 
Note that for the ground state given by (\ref{psi0}), $Z=0$.

The functions $\tilde\phi(x)$ and $\tilde\chi(x)$ span a two-dimensional subspace of the continuous-variable Hilbert space. To calculate the reduced spin density matrix, it is convenient to introduce an orthonormal set of two functions
 \begin{eqnarray}
 f_1(x)={1\over\sqrt{2(1+|Z|)}}\left(\tilde\phi(x)+e^{-i\alpha}\tilde\chi(x)\right), \\
 f_2(x)={1\over\sqrt{2(1-|Z|)}}\left(\tilde\phi(x)-e^{-i\alpha}\tilde\chi(x)\right),
 \end{eqnarray}
that satisfy $\int_{-\infty}^{\infty}dx f^*_i(x)f_j(x)=\delta_{ij}$. Inverse transformation reads
\begin{eqnarray}
\tilde\phi(x)={1\over\sqrt 2}\left(\sqrt{1+|Z|}f_1(x)+\sqrt{1-|Z|}f_2(x)\right), \\
\tilde\chi(x)={e^{i\alpha}\over\sqrt 2}\left(\sqrt{1+|Z|}f_1(x)-\sqrt{1-|Z|}f_2(x)\right)
\end{eqnarray}
Substituting it into (\ref{psi1}) we find
\begin{eqnarray} \nonumber
|\psi\rangle=\sqrt{1+|Z|\over 2}\left( 
a|\uparrow\rangle +e^{i \alpha}b|\downarrow\rangle \right)f_1(x)+ \\ 
\sqrt{1-|Z|\over 2}\left( 
a|\uparrow\rangle -e^{i \alpha}b|\downarrow\rangle \right)f_2(x).
\end{eqnarray}

By tracing the pure density matrix $|\psi\rangle\langle\psi|$ over the continuous-variable degrees of freedom, we obtain the reduced density matrix for the spin subsystem 
\begin{eqnarray}
\hat\rho=a^2|\uparrow\rangle \langle\uparrow|+ab|Z|e^{-i\alpha}
|\uparrow\rangle \langle\downarrow|+\\
ab|Z|e^{i\alpha}|\downarrow\rangle \langle\uparrow|+
b^2|\downarrow\rangle \langle\downarrow|.
\end{eqnarray}
The eigenvalues of the reduced density matrix are
\begin{eqnarray}\nonumber
\lambda_{1,2}={1\over 2}\left(1\pm\sqrt{1-4a^2b^2(1-|Z|^2)}\right)=\\ \label{Lambda12Gen}
={1\over 2}\left(1\pm\sqrt{{m^2c^4\over E^2}+\left(1-{m^2c^4\over E^2}\right)|Z|^2}\right),
\end{eqnarray}
where $E=c\sqrt{m^2c^2+\epsilon}\ge mc^2$.

The entanglement entropy is given by
\begin{eqnarray}
S=-\lambda_1 \log_2 \lambda_1-\lambda_2 \log_2 \lambda_2.
\end{eqnarray}

We see that for the ground-state energy $E=mc^2$, the eigenvalues are
$\lambda_{1,2}=1,0$. As a result, the entanglement entropy vanishes, $S=0$.

In the limit $E\to\infty$,
\begin{eqnarray}
\lambda_{1,2}={1\over 2}\left(1\pm |Z|\right),
\end{eqnarray}
and the entanglement entropy is determined solely by the overlap integral.

Note that SUSY also makes it possible to rewrite the overlap integral as
\begin{eqnarray}
Z=\int_{-\infty}^{\infty}dx \tilde\phi(x){1\over\sqrt\epsilon}\hat B\tilde\phi(x)=\\
={1\over\sqrt\epsilon}\int_{-\infty}^{\infty}dx \tilde\phi(x)(\hat p-iW(x))\tilde\phi(x)=
{-i\over\sqrt\epsilon}\int_{-\infty}^{\infty}dx |\tilde\phi(x)|^2W(x),
\end{eqnarray}
where $\epsilon\ne 0$.
Here we have used the fact that $\tilde\phi(x)$ can be chosen to be a real function up to an overall constant phase, which can be set to zero. Consequently, the expectation value of the momentum operator in this state vanishes. Thus, the overlap integral is determined by the expectation value of the superpotential in the state $\tilde\phi(x)$.

Similarly we find
\begin{eqnarray}
Z^*=\int_{-\infty}^{\infty}dx \tilde\chi^*(x){1\over\sqrt\epsilon}
\hat B^+\tilde\chi(x)=
{i\over\sqrt\epsilon}\int_{-\infty}^{\infty}dx |\tilde\chi(x)|^2W(x).
\end{eqnarray}
We see that the overlap integral can also be expressed as the expectation value of the superpotential in the state $\tilde\chi(x)$.

Comparing $Z$ and $Z^*$, we find that for $\epsilon\ne 0$ ($E\ne mc^2$),
\begin{eqnarray}
\int_{-\infty}^{\infty}dx |\tilde\phi(x)|^2W(x))=\int_{-\infty}^{\infty}dx |\tilde\chi(x)|^2W(x).
\end{eqnarray}

In general, the dependence of $Z$ on energy is determined by the superpotential $W(x)$. In the case of an odd superpotential, $W(-x)=-W(x)$, the overlap integral vanishes. In this case, $\tilde\phi(x)$ and $\tilde\chi(x)$ have opposite parity under the reflection $x \to -x$. As a result, $\tilde\phi(x)$ and $\tilde\chi(x)$ are orthogonal, and $Z=0$ for all energies.

In this case, the eigenvalues read
\begin{eqnarray}
\lambda_{1,2}={1\over 2}\left(1\pm{mc^2\over E}\right).
\end{eqnarray}

The entanglement entropy then increases monotonically with energy $E$ and reaches its maximal value $S=1$ in the limit $E\to \infty$.

Let us analyze the dependence of the entanglement on the speed of light. In the limit $c \to \infty$, Eq. (\ref{Lambda12Gen}) yields $\lambda_1 = 1$, $\lambda_2 = 0$, and therefore $S = 0$. Thus, in the nonrelativistic limit, the entanglement between spin and continuous variables disappears. This suggests that the entanglement has a purely relativistic origin.

\section{Conclusion}

In this paper, we have studied the entanglement between the spin degree of freedom and a continuous variable in a (1+1)-dimensional generalized Dirac oscillator. This system exhibits supersymmetric properties, which allow us to determine the entanglement entropy purely algebraically, without explicitly solving the eigenvalue equation. SUSY is crucial for obtaining these results.

We have shown that the entanglement entropy depends on the energy and on the overlap integral between the normalized large and small components of the eigenstates of the generalized Dirac oscillator (i.e., the eigenstates of the SUSY partner Hamiltonians). The overlap integral can be expressed as the expectation value of the superpotential in either the normalized small or large component of the Dirac eigenstates.

We further show that the entanglement has a purely relativistic origin and that, in the nonrelativistic limit, the entanglement between spin and continuous variables disappears. The ground state with energy $E=mc^2$ exhibits no entanglement. In the limit of large energies, the entanglement entropy is determined solely by the overlap integral.

If the system possesses an additional spatial reflection symmetry, namely if the superpotential is an odd function, $W(-x)=-W(x)$, then the overlap integral vanishes, $Z=0$. In this case, the entanglement entropy depends only on the energy of the eigenstates and is a monotonically increasing function of energy. The maximal possible value of the entanglement entropy ($S=1$) is reached in the limit of infinite energy.

\section*{Acknowledgments}
The authors are grateful to Dr. M. Samar for valuable discussions and useful comments.
This work was supported by Project 2025.07/0108 from the National Research
Foundation of Ukraine.

\end{document}